%% file: MPS_12012018.tex
\documentclass[aps,reprint,prl]{revtex4-1}
\usepackage{graphicx,amsmath}
\input{kleinPreamble}
\newcommand{\beq} {\begin{equation}}
\newcommand{\eeq} {\end{equation}}
\newcommand{\bea} {\begin{eqnarray}}
\newcommand{\eea} {\end{eqnarray}}
\renewcommand{\ef}{E_F}
\begin{document}

\title{
  Multiple pairing states and temperature-dependent gap anisotropy for superconductivity near a nematic quantum-critical point}

\author{Avraham Klein}
\author{Yi-Ming Wu}
\author{Andrey Chubukov}
\affiliation{School of Physics and Astronomy, University of Minnesota, Minneapolis. MN 55455}

\begin{abstract}
  Superconductivity in many strongly correlated materials
  appears in proximity to a density-wave or nematic order and is believed to be mediated by quantum-critical (QC) fluctuations of the corresponding order parameter.
  We argue that fingerprints of QC pairing can be extracted from the  angular dependence of the gap $\Delta (\theta)$. We consider pairing by QC nematic fluctuations and show
  that there exist multiple pairing instabilities within the same symmetry ($s-$wave in our case),  with closely spaced transition temperatures $T_{c,n}$. The corresponding $\Delta_n (\theta)$ change sign $8n$ times along the FS. Only the solution with the highest $T_{c,0} =T_c$ develops, but other gap components are induced below $T_c$ and get resonantly enhanced below $T_{c,n}$.  This gives rise to strong variation of the angular dependence of the gap below $T_c$. The effect gets much weaker away from a quantum-critical point.
\end{abstract}
\maketitle

\paragraph*{Introduction --}

In  many correlated materials superconductivity (SC) has been observed in proximity to either a density-wave or a nematic order~\cite{Fradkin2015}.
A SC dome in the phase diagram of these materials is often considered an indication
of pairing  mediated by order parameter fluctuations in a quantum critical (QC) regime.
However, SC dome can appear for other reasons as well, e.g., due to the ``fight" for the Fermi surface (FS) between SC and density-wave orders, even when each is described  within BCS/mean-field theory. The question we address here is whether there are unique features of SC mediated by QC fluctuations encoded in the SC gap structure on the Fermi surface,
$\Delta (\theta, \omega)$. Previous studies have focused on signatures of QC pairing in the frequency/temperature dependence of the gap function \cite{McMillan1965,Campuzano2004,Scalapino2012}.
Such emphasis stems from the understanding that the pairing kernel in the QC regime is a singular function of frequency. In contrast, the angular dependence of the gap along the FS was assumed to be set either by the non-s-wave pairing symmetry (e.g., $d-$wave in the cuprates ~\cite{Shen1993,Hashimoto2014}), or, for s-wave,  by some  material specific non-singular angular dependencies of interactions and  band structures, as in the Fe-based superconductors
\cite{Xu2016,Sprau2017,Kushnirenko2018,Hashimoto2018,Allan2012}.
In either case the angular variation of the gap $\Delta (\theta)$ was expected to be set at $T_c$ and not vary strongly in the SC state.

Contrary to this view, we argue in this Letter that  QC pairing can give rise to a strong temperature evolution of $\Delta (\theta)$ below $T_c$ at any given $\omega$.  Specifically, we argue that this is the case for $s-$wave SC near a nematic transition in 2D \cite{Hashimoto2018,Smylie2018,Sprau2017,Kuo2016,Xu2016,Fernandes2014,Chu2012,Fernandes2012,Chu2010,Ronning2017,Thorsmolle2016,Zhang}. We show that the mechanism driving this evolution is the existence of multiple $s-$wave pairing states with orthogonal gap functions $\Delta_n(\theta)$ and closely spaced transition temperatures $T_{c,n}$. These solutions yield $\Delta_n(\theta)$  of the same symmetry, but with a different number ($=8n$) of sign changes of the gap along the FS.
Only the solution with the highest $T_{c,0} = T_c$ actually develops, as it has the largest condensation energy. Still, because other $\Delta_n$ are nonlinearly coupled to $\Delta_0$ in the Free energy, the physical $\Delta (\theta)$ coincides with $\Delta_0 (\theta)$ only near $T_c$, while at smaller $T$ other $\Delta_n (\theta)$ contribute to the gap structure. By itself, this effect is an expected one because in a lattice system there is an infinite number of orthogonal gap components in an $s-$wave ($A_{1g}$) channel, and all these components are induced below $T_c$. However, these induced components are usually quite small. In our case, other components could by themselves develop at $T_{c,n}$, and their contributions to $\Delta(\theta)$ are resonantly enhanced. This leads to strong evolution of the gap structure below $T_c$.

We argue that multiple closely spaced solutions for $T_{c,n}$ exist only in proximity to the
nematic QCP.  Away from a QCP, other solutions shift to smaller $T$ and progressively disappear as the nematic correlation length
gets smaller. An experimental observation of a strong variation of the shape of $\Delta(\theta)$ below $T_c$ would then be conclusive evidence for QC pairing.

\begin{figure}[t!]
  \centering
  \includegraphics[trim=50 150 50 150, clip,width=0.7\hsize]{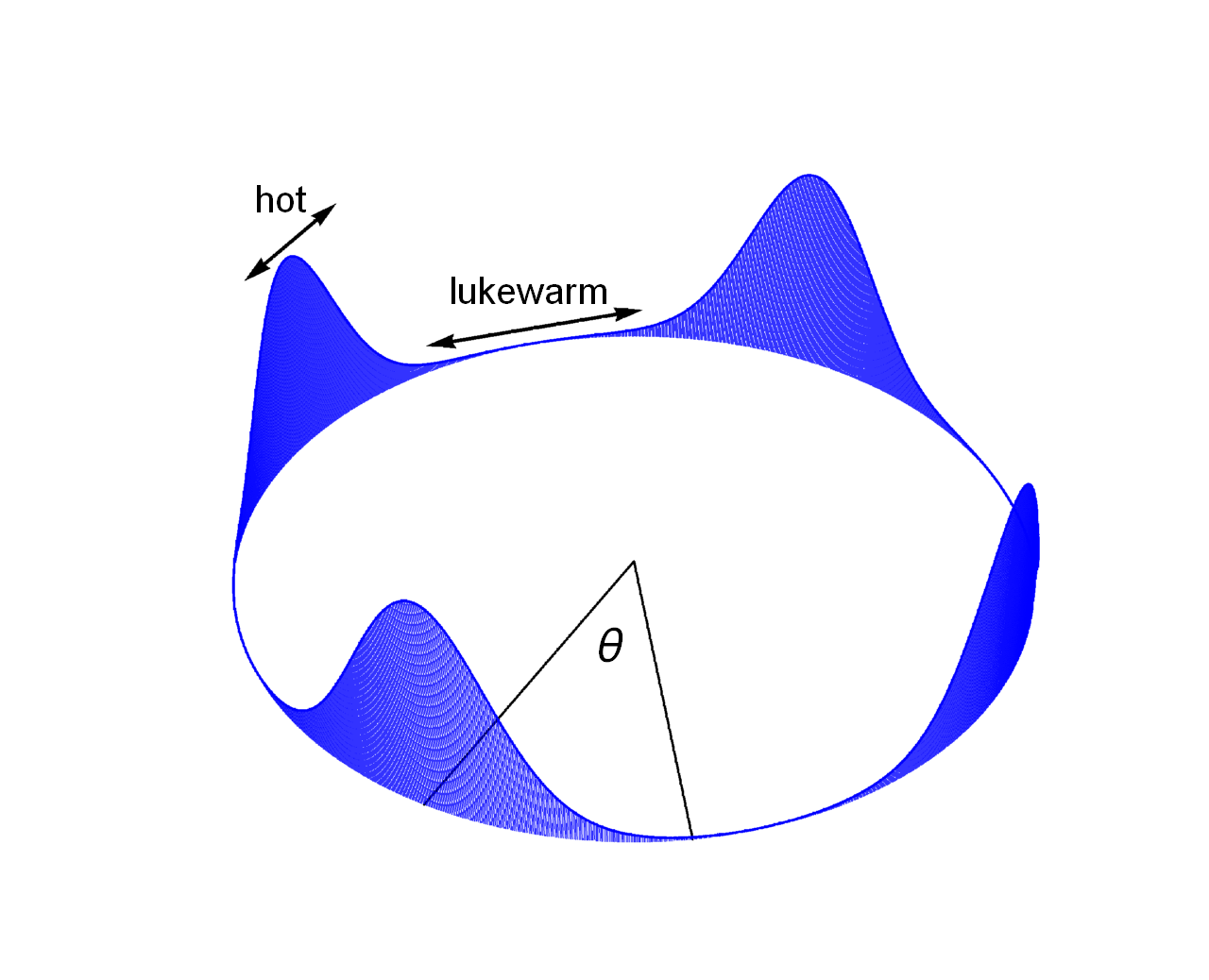}
  \caption{The
  s-wave gap function around the Fermi surface
   for a superconducting state near a nematic QCP.
  The gap
   is sharply peaked in four hot regions surrounding the points
    $\theta = n \pi/2 $, where the interaction mediated by QC nematic fluctuations is the strongest, and is strongly suppressed
    around
     $\theta = \pi/4+n \pi/2$, where the interaction is the weakest.}
  \label{fig:gap-form}
\end{figure}

\begin{figure}
  \centering
  \includegraphics[width=0.75\hsize]{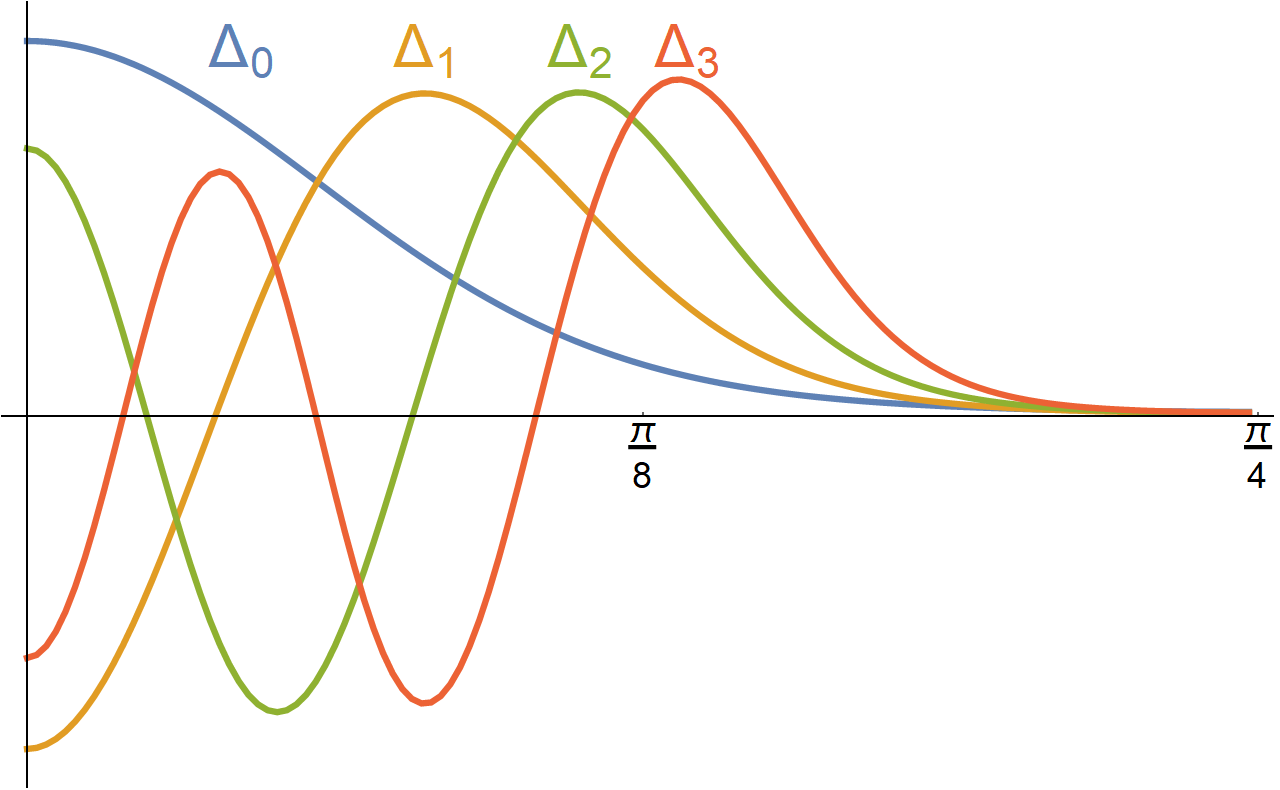}
  \caption{Orthogonal solutions of the linearized gap equation. We used $\lambda = 0.025$ to produce this figure.}
  \label{fig:solutions}
\end{figure}

We begin with a qualitative description of our results and the physics behind the appearance of multiple pairing states. We consider $s-$wave superconductivity of 2D fermions, minimally coupled to order parameter fluctuations near a nematic QCP.
~\cite{Bonesteel1996,Altshuler1994,Nayak1994,Nayak1994a,Rech2006,Lee2009, Maslov2010,Fradkin2010,Metlitski2010a,Lee2018,Raghu2015,Metlitski2015}.
The strength of fermion-boson interaction is described by a dimensionless coupling $\lambda$ (defined below). We consider a metal far from a Mott transition and treat $\lambda$  as a small parameter.  The nematic order develops at $q = 0$, so the pairing interaction is peaked at zero momentum transfer and can involve fermions at any angle $\theta$ on the FS. However, this interaction also depends on $\theta$ via the square of the nematic form-factor $f^2(\theta)= \cos^2{2\theta}$. As a result, the pairing gap is the largest in `hot' regions near $\theta = \frac{\pi}{2}m$, and is smaller in 'cold' regions near
$\theta = \frac{\pi}{4}+m\frac{\pi}{2}$ (see Fig. \ref{fig:gap-form}) \cite{Lederer2015,Klein2018c}. We show that  the width of the hot region increases with decreasing $T<T_c$ due to induced contributions from other $\Delta_n (\theta)$. This is a non-perturbative effect, arising uniquely from the proximity to
a nematic QCP, at which the induced contributions are amplified because
other pairing states would develop on their own at $T_n$ only slightly below $T_c$,

At the QCP, the scale of bosonic dynamics (the Landau damping) is set by the same interaction which gives rise to the pairing. As a result, strong pairing occurs for fermions within a small angular separation of order $\delta\theta \sim \lambda$. To leading order in $\lambda$, the gap equation then  becomes local, and allows a continuous set of solutions
$T_c(\theta)$
with
$\Delta (\theta') \propto \delta (\theta - \theta')$.
The physical $T_c$ is the highest temperature in the set, and it corresponds to $\theta=\pi/2 m$. The actual gap structure is determined at the next approximation, when one properly accounts for
weaker interactions at angle transfers well above $\delta \theta$. The result is that in each
octant, e.g. at $0 < |\theta| < \pi/4$,  the actual gap remains of order $\Delta (\theta =0)$ for
$|\theta|  < \theta_h \sim \lambda^{1/3}$ ($\delta \theta \ll \theta_h \ll 1$) and rapidly drops as $(\theta_h/\theta)^4$ at larger $\theta$. This can be interpreted as if  a Cooper pair is `trapped' in a potential well within the range $\theta_h$. The  correlation function between fermions in a pair (the gap function) can vary
inside the trap at the cost of a small kinetic energy $\sim \lambda/\lambda^{1/3} = \lambda^{2/3}$.
This near-freedom implies that even
 after accounting for
weaker interactions there still exists a series of pairing states with gap functions $\Delta_n$ and onset temperatures differing from $T_{c,0}$ by multiples of the kinetic energy cost
\begin{equation}
  \label{eq:varepsilon-intro}
  T_{c,n} = T_{c,0} \left (1 - O(n \lambda)^{2/3}\right), ~~ T_c = T_{c,n=0}
\end{equation}
Fig. \ref{fig:solutions} depicts several such solutions. They are all fourfold periodic,
i.e., $s-$wave, but have $n$ nodes in the first
octant ($8n$ total along the FS). They can be approximated analytically by solutions of the Airy equation,  a natural result of linearizing the `trapping' gap potential~\cite{nematicSupp2018}. Solutions exist up to $n_{max} \sim 1/\lambda \gg 1$.

\begin{figure}[t!]
  \centering
 \includegraphics[width=\hsize]{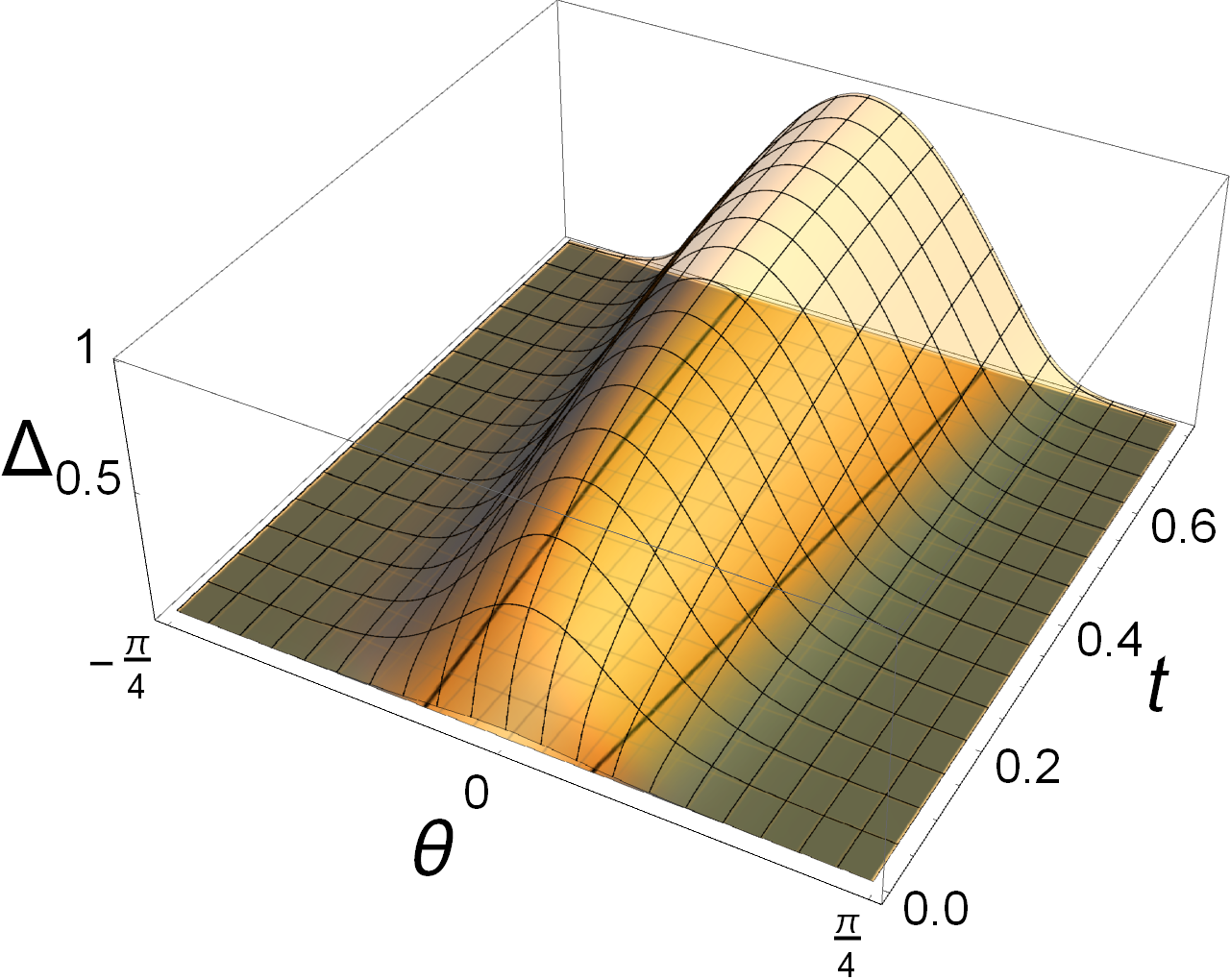}\llap{\makebox[\hsize][l]{\raisebox{0.55\hsize}{\frame{\includegraphics[width=0.42\hsize]{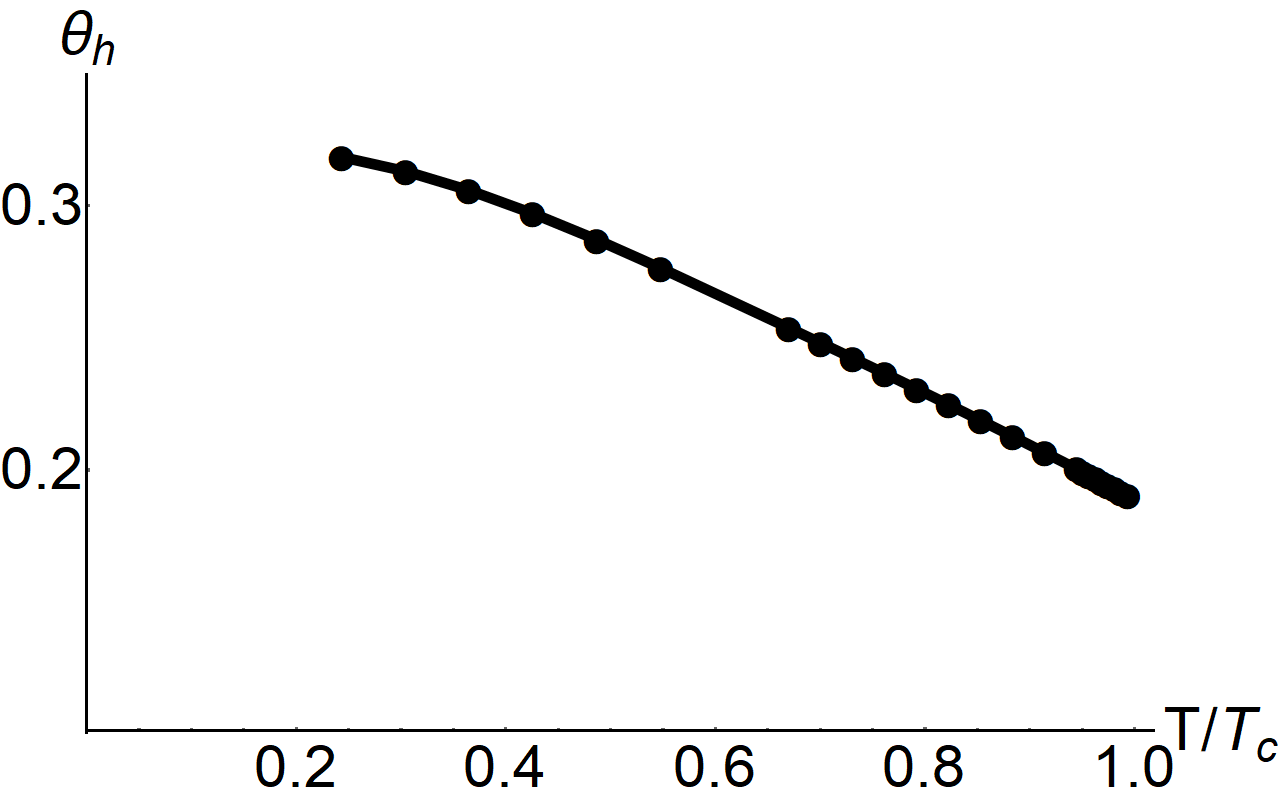}}}}}
  \caption{Color online. Evolution of the gap anisotropy with temperature. The 3D plot depicts the pairing gap $\Delta(t,\theta)$ at $\omega_n = \pi T$, as a function of the angle along the FS, counted from $\theta=0$, and reduced temperature $t = (T_c - T)/T_c$. The gap function has been obtained by numerically solving the full nonlinear Eliashberg gap equation at a nematic QCP with coupling
  $\lambda = 0.03$.
  $\Delta$ has been normalized to its maximum value at $T = 0$. The inset shows the temperature dependent width of the gap $\theta_h(t) \propto \int \Delta (t, \theta) d \theta$.  }
  \label{fig:gap-evolution}
\end{figure}

At $T < T_c$, the physical gap is a superposition of $\Delta_0$ and $\Delta_{n>0}$, which are induced by $\Delta_0$.
This superposition causes destructive interference in the trap region, where oscillations occur, and constructive interference outside it. As a result, the gap width increases strongly with decreasing temperature, as more and more oscillating $\Delta_{n>0}$ are superimposed on the
non-oscillating $\Delta_0$. Fig. \ref{fig:gap-evolution} shows a numerical solution of the nonlinear gap equation and strong evolution of gap width with decreasing temperature.  In the rest of this paper we substantiate our qualitative arguments by solving the non-linear gap equation analytically and numerically.

\paragraph*{Model and gap equation --}
We consider 2D fermions with Fermi energy $\ef$, minimally coupled to a nematic order parameter field $\phi(q)$ by
\begin{equation}
  \label{eq:H-int}
  H_I = g\sum_{\k,\q,\sigma}\phi(\q)f(\k)\psi^{\dagger}_\sigma\left(\k+\frac{\q}{2}\right)\psi_\sigma\left(\k-\frac{\q}{2}\right),
\end{equation}
where $f(\k)$ is a form-factor, which has $d-$wave symmetry with respect to $C_4$ lattice rotations.  We assume the static susceptibility of
the $\phi$-field is peaked at $q=0$: $\chi^{-1} (q) = \chi_0^{-1}(\xi_0^{-2} + q^2)$.
We define the
dimensionless coupling to be
$\lambda = g^2 \chi_0/4\ef$.
We assume for simplicity a circular FS, but our results are readily generalized to other $C_4$-symmetric FSs. Because relevant fermions are near the FS, we can approximate $f(\k)$ by $f(\theta)=\cos2\theta$ (see Fig. \ref{fig:gap-form}). A bosonic excitation with momentum $\q$ connects two fermions on the FS with angles $\theta,\theta+\phi$ such that $\q = 2\k_F\sin|\phi/2| \approx \kf |\phi|$.
The effective pairing interaction is then the function of both $\phi$ (via $\chi (q)$) and $\theta$ (via the form-factor $f(\theta)$). This interaction modifies both the bosonic and fermionic dynamics. Fermionic self-energy scales as $\Sigma (\omega_m) \sim (\lambda^2 E_F)^{1/3} \omega^{2/3}_m$ modulo logarithmic corrections from higher-order planar diagrams~ ~\cite{Lee2009,Metlitski2010a,Holder2015,Klein2018},
  which we neglect here, and singular contributions from thermal fluctuations at $T>0$. The latter cancels out between self-energy and the pairing vertex~\cite{Millis1988,Abanov2008,Wang2016} and we eliminate it in $\Sigma (\omega_m)$ and in the gap equation (\ref{eq:gap-eliashberg-1}) below.
The  bosonic self-energy gives the Landau damping term,  and with it the bosonic susceptibility  becomes
\begin{equation}
  \label{eq:D-def-ang}
  \chi(\theta,\phi,\W)^{-1} =
 \frac{k^2_F}{\chi_0}
 \left(|\phi|^2
    +
 \frac{\lambda}{\pi E_F}
   f^2(\theta)\left|\frac{\W}{\phi}\right|\right),
\end{equation}
We assume and verify that fermionic and bosonic frequencies $\w$ and $\Omega$, relevant to the pairing, are of order $\lambda^2 E_F \ll E_F$. Then  a typical fermionic momentum transfer transverse to the FS is of order $\Sigma (\omega)/v_F \sim (\lambda^2 E_F)^{1/3} \w^{2/3}/v_F \sim \lambda^2 k_F$, while a typical bosonic momentum in Eq. \eqref{eq:D-def-ang} of order $\kf \phi \sim k_F (\lambda \w/E_F)^{1/3} \sim \lambda k_F$, i.e., it is larger by $1/\lambda$. In this situation,  one can integrate out transverse momenta in the gap equation and obtain an equation for the gap function $\Delta (\theta, \omega_n)$ on the FS \cite{Klein2018c}:
\begin{align}
  \label{eq:gap-eliashberg-1}
  Z(\theta,\w_n)\Delta(\theta,\w_n) &=
  T\sum_{n' \neq n}\int \frac{d\phi}{\tp} \frac{\Delta(\theta+\phi,\w_{n'})} {|\w_{n'}|} S_{n-n'}(\theta, \phi) \nn\\
  S_{n-n'}(\theta, \phi) &= f^2\left(\theta+\frac{\phi}{2}\right) D(\theta,\phi,\w_{n'}-\w_n).
\end{align}
Here $D (\theta, \phi, \Omega) = \lambda k^2_F \chi (\theta, \phi, \Omega)/\chi_0$ is the normalized susceptibility and $Z(\theta,\w_n) = (1+\Sigma/\w_n)^{-1}$ is the quasiparticle residue.

Eq. (\ref{eq:gap-eliashberg-1}) allows solutions with different gap symmetry.  Earlier works found~\cite{Lederer2015,Klein2018c} that an $s-$wave SC state has larger condensation energy,
so we focus on $s-$wave gap symmetry. Because $D(\theta,\phi,\w_{n'}-\w_n)$ is strongly peaked at $\phi = 0$, to first approximation one may set $\phi = 0$ in $\Delta(\theta+\phi,\w_m)f^2\left(\theta+\frac{\phi}{2}\right)$.  This yields a local gap equation for a frequency dependent gap $\Delta (\theta, n) = \Delta_\theta (n)$ with $\theta$ acting as a parameter~\cite{Bonesteel1996,Moon2010,Wang2016,Lee2018,Wu2018a}
At the QCP we have
\begin{equation}
  \label{eq:Delta-strong}
  \Delta_\theta (n) \approx
  \frac{1}{3^{3/2} \pi Z_\theta (n)}
    \sum_{n'}\frac{\Delta_\theta (n')}{|n'+1/2|}\frac{1}{\left(
        \frac{2 T_c} {\lambda^2E_F f^4 (\theta)}|n'-n|\right)^{1/3}}.
\end{equation}
Eq. \eqref{eq:Delta-strong} has a continuous set of solutions
$\Delta_{\nu}(\theta, n)\propto \delta(\theta - \nu)$ with $ T_{c} (\nu) \sim \lambda^2 \ef
 f^2(\nu)$. The maximum $T_c (0) \approx
0.022 \lambda^2 E_F
 f^4 (0)$ corresponds to $\nu =0$.
To determine the actual structure of $\Delta (\theta,n)$ and the correct number of solutions, we need to go beyond the leading approximation and keep the dependence on $\phi$ in the numerator in (\ref{eq:gap-eliashberg-1}).
The problem is analytically tractable if we utilize the fact that
typical $\omega_{n,n'} \sim T$, i.e., typical $n, n' = O(1)$ and typical  $Z = O(1)$, and simplify the gap equation by neglecting the frequency dependence of $\Delta$ and setting $Z=1$. In this approximation, Eq. \eqref{eq:gap-eliashberg-1} becomes an effective 1D integral equation over the angle. We expand in small angles near $\theta =0$  and for temperatures near the highest $T_c$ and obtain \cite{nematicSupp2018}
\begin{equation}
  \label{eq:gap-dimen-full}
   \eta (t) \Delta({\bar \theta}) \approx \bar\theta^2 \Delta({\bar \theta}) - \int_{-\infty}^{\infty}\frac {d{\bar \phi}}{\pi}\frac{\Delta({\bar \theta}+{\bar \phi})-\Delta({\bar \theta})}{{\bar \phi}^2}
\end{equation}
Here $t = (T_c-T)/T_c$, $\theta_h \sim \lambda^{1/3}$, $\eta (t) \sim t/\theta_h^2$, and $({\bar \theta},{\bar \phi}) = \theta_h (\theta,\phi)$ are rescaled angular variables.
Because Eq. (\ref{eq:gap-dimen-full}) has no parameters,  $\theta_h$ sets the width of the gap function $\Delta (\theta)$  in actual $\theta$. Transforming to a Fourier representation $\Delta({\bar \theta}) = \tpp^{-1}\int d\xi \exp(i \xi {\bar \theta}) \Delta({\bar \theta})$, we find
\newcommand{\Ai}{\mbox{Ai}}
\begin{equation}
  \label{eq:gap-Airy}
  \eta (t) \Delta (\xi) = -\pd_\xi^2\Delta (\xi) + |\xi| \Delta (\xi).
\end{equation}
This is the Airy equation
(for $\xi > 0$).
It has solutions
\begin{equation}
  \label{eq:Airy-sols}
\eta_n = \xi_n, ~\Delta_n(\xi)=\Ai(|\xi| - \xi_n),
\end{equation}
where $\xi_n$ is the $n^{\text{th}}$ zero
of the Airy function $\Ai(|x|)$
for odd $\Delta(\theta)$ and of its derivative $\Ai'(x)$ for even $\Delta(\theta)$.
As we described in the Introduction, the appearance of the Airy equation is 
 attributed to the fact that the Cooper pair can oscillate in the `trapping potential' set by $\theta_h$.

The smallest eigenvalue $\eta_n$ is for the even solution with $n=0$. It yields a  non-oscillating gap $\Delta_0 (\theta)$, peaked at $\theta = 0$, with the width $O(\theta_h)$.  The corresponding $T_{c,0}$ is the actual $T_c$ for the pairing at a nematic QCP. It differs by a
 numerical factor of order $\langle f^4 (\theta)/f^4(0)\rangle \sim (1 - O(\lambda^{2/3}))^4$
 from $T_c (0)$ from  Eq. (\ref{eq:Delta-strong})
 (for
 $\lambda =0.03,
 ~T_{c,0} \approx 0.75 T_c (0)$).
   For other solutions, $\Delta_n (\theta)$ changes sign $n$ times in the first
octant.  We focus on even $\Delta_n (\theta)$ as only even solutions will be generated by $\Delta_0 (\theta)$ at $T < T_{c,0}$. For large
negative $\xi$,
$\Ai(\xi)\sim \sin\left( \frac{2}{3}|\xi|^{3/2}+\frac{\pi}{4}\right)/\sqrt{\pi}|\xi|^{1/4}$,
and
 $T_{c,n} = T_{c,0} \left( 1 - O(n \lambda)^{2/3}\right)$,
 up to $n_{max} \sim 1/\lambda \gg 1$.
Fig. \ref{fig:solutions} depicts the first few even solutions.

\paragraph*{Nonlinear gap equation and evolution of $\theta_h$ --}

At $T$ only slightly below  $T_{c,0} = T_c$,  the angular dependence of the pairing gap coincides with $\Delta_0 (\theta)$, only the overall gap magnitude increases with decreasing $T$.  We now show that the angle dependence of $\Delta (\theta)$ drastically changes at smaller $T$ because solutions with $\Delta_{n>0}$ strongly influence the evolution of the gap width. The reason for this is that  below $T_c$, $\Delta_0 (\theta)$ induces the orthogonal gap components with $n>0$, and the $n^{\text th}$  component gets near-resonantly enhanced at $t \sim n \lambda^{2/3}$. These additional gap components interfere destructively in the region $|\theta| < \theta_h$, where they oscillate, but constructively for $\theta > \theta_h$. To study this behavior we introduce Ginzburg-Landau Free energy and include
only those quartic terms that couple
$\Delta_0$  and $\Delta_n$:
\begin{align}
  \label{eq:free-energy}
  F &\approx -\frac{1}{2}\sum_n(t-t_n)\bar{\Delta}^2_n + \sum_n\frac{a_{n}}{4}\bar{\Delta}_n^4 \nn\\
    &\qquad
 + \sum_{n>0}\left[a_{n1} \bar\Delta_n^3\bar\Delta_0 + a_{n2}  \bar\Delta_0^2\bar\Delta_n^2 +a_{n3}\bar\Delta_0^3\bar\Delta_n
 \right].
\end{align}
Here, $\bar\Delta_n = \bar\Delta_n (t)$ represent the coefficients in the gap function,
$\Delta(t,\theta) = \bar\Delta_0\Delta_0(\theta) + \bar\Delta_1\Delta_1(\theta) + \cdots$, where
$\Delta_n (\theta)$ are  orthonormal eigenfunctions of the linearized gap equation, $t_n = (T_{c,0}-T_{c,n})/T_{c,0}$ is a reduced onset temperature for the $n^{\text th}$ solution, and
$a_{nk} = \int d\theta \Delta_n^{4-k}(\theta)\Delta_0^k(\theta)$.
 At small $\theta$, $\Delta_n (\theta) \propto (-1)^n$,  at large $\theta \gg \theta_h$ all $\Delta_n (\theta)$, including $n=0$,  decay as $1/\theta^4$.
 We verified numerically that $a_{n2} \gg |a_{n1}|, |a_{n3}|$ for $n > 1$, and that $a_{n3} \propto (-1)^n$. Solving the saddle point equation, we find that immediately below $T_c = T_{c,0}$, when $t -t_0 =t \ll 1$ and $t-t_n \approx -t_n$, we have $\bar\Delta_0^2 \approx t /a_0$ and
${\bar \Delta}_n \approx - {\bar \Delta}_0 (a_{n3}/a_0) t/t_n$, i.e., $|\bar\Delta_n| \ll {\bar \Delta}_0$.
 In the opposite case when $t \geq  t_n$, we have
 $\bar\Delta_n \approx (-1)^{n+1} \bar\Delta_0\sqrt{(a_0(t - t_n) - 2 a_{n2} t)/ t a_n}$,
  i.e., $|{\bar \Delta}_n|$  is of the same order as ${\bar \Delta}_0$.
  Because $\Delta_n(\theta) \propto (-1)^n \Delta_0$ at small $\theta$ and ${\bar \Delta}_n \propto (-1)^{n+1} {\bar \Delta}_0$,
  ${\bar \Delta}_n \Delta_n(\theta)$ for all $n >0$ have opposite sign compared to
  ${\bar \Delta}_n \Delta_n(\theta\approx 0)$. The original and the induced gap components then interfere destructively, and the total
  $\Delta (\theta\approx 0)$ decreases with decreasing $T$. On the other hand, at large $\theta \gg \theta_h$, $\Delta_n (\theta) \propto \Delta_0$, hence  ${\bar \Delta}_n \Delta_n(\theta)$ oscillates in sign between even and odd $n$, and the sign of the largest  ${\bar \Delta}_1 \Delta_1(\theta)$ is the same as of  ${\bar \Delta}_0 \Delta_0(\theta)$. Then the original and the induced gap components interfere mostly constructively. As the consequence, the  effective 'width' of the  gap, $\theta_h (t)  \propto \int \Delta (t,\theta) d \theta$ should get larger with decreasing $T$.
F or $t \sim t_N$, when $N \gg 1$ states are hybridized, we find, using $N \propto t^{3/2}$,
\begin{equation}
  \label{eq:theta-p-evol}
  \frac{\theta_h(t) - \theta_h(0)}{\theta_h(0)} = a t + b t^{5/4}
\end{equation}
where $a t$  and $b t^{5/4}$ are the contributions from states with $t_n > t$ and $t_n < t$, respectively.
Fig.~\ref{fig:gap-evolution} displays the numerical solution of the nonlinear gap equation. We see  that the width of the ``hot" region indeed strongly increases with decreasing temperature as $t^\alpha$, and $\alpha \approx 1$, consistent with Eq. (\ref{eq:theta-p-evol}).
We emphasize that a good numerical agreement with our predictions holds
even when the numerical value of $\theta_h \sim  0.3$, i.e. a hot region is fairly broad.
\begin{figure}[t!]
  \centering
  \includegraphics[width=0.8\hsize]{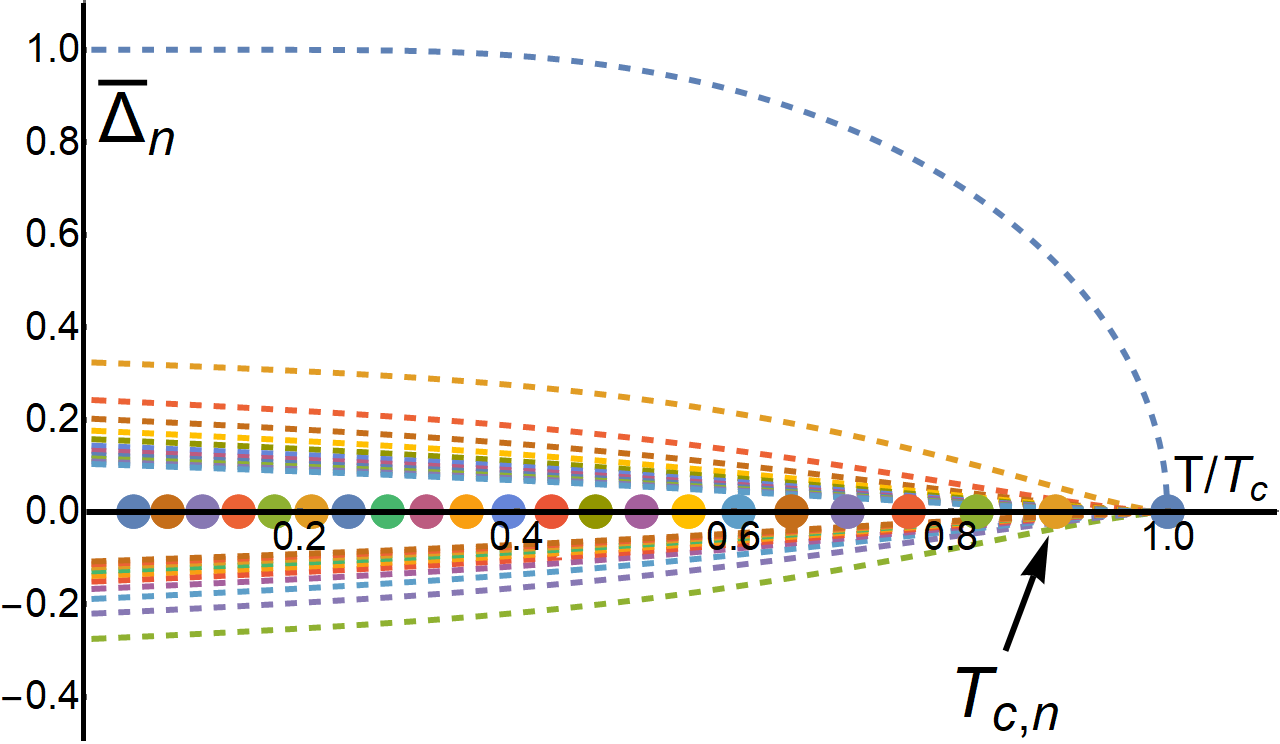}
  \caption{
  The temperature evolution of the induced  ${\bar \Delta}_n$ of $\Delta(\theta) = \bar\Delta_0\Delta_0(\theta) + \bar\Delta_1\Delta_1(\theta) + \cdots$, where
$\Delta_n (\theta)$ are orthonormal eigenfunctions of the linearized gap equation.
The contribution of $\bar\Delta_n\Delta_n(\theta)$ to $\Delta_n (\theta)$
 is  resonantly enhanced for $T<T_{c,n}$.  Observe that the sign of ${\bar \Delta}_n$ oscillates between even and odd $n$:
 ${\bar \Delta}_n \propto (-1)^{n+1} {\bar \Delta}_0$. Because at small $\theta$, $\Delta_n (\theta) \propto (-1)^n \Delta_0 (\theta)$,
  the original ($n=0$) and the induced ($n>0$) gap components
  interfere  deconstructively.}
  \label{fig:d-n-sketch}
\end{figure}
\paragraph*{Away from the QCP --} Our results are readily generalized \cite{nematicSupp2018} to a system at a finite distance from the QCP, when the correlation length $\xi$ for nematic fluctuations is large but finite.
For $(\kf\xi)^{-1} \ll \lambda$ we find that
the effect of $\xi$ is a uniform reduction of all onset temperatures. For larger $\lambda \ll (\kf\xi)^{-1} \ll \lambda^{3/5}$, the onset temperatures become BCS like, $T_{c,0} \sim \exp(-(\kf\xi)^{-1}/\lambda)$, and $ T_{c,n} = T_{c,0}\left(1 - n^{2/3}(\bar k_F\xi)^{-5/3}/\lambda \right)$, with $\bar k_F \sim \kf$.
As a result, fewer
gap components get resonantly enhanced, and 
 the temperature 
variation of the shape of $\Delta (\theta)$
below $T_c$
 weakens and becomes
undetectable at
$(\kf \xi)^{-1} \sim \lambda^{3/5}$.
Consequently, 
 this variation 
  is
a fingerprint of QC pairing.

\paragraph*{Summary and discussion --}
We analyzed s-wave superconductivity near a nematic transition in 2D.
We showed that the  gap function, induced by  quantum-critical nematic fluctuations, is peaked in hot regions, and  has a strong and distinctive temperature evolution within the superconducting state. The peak width grows with decreasing temperature.  The source of this evolution is the existence of multiple solutions for the pairing gap $\Delta_n (\theta)$ with closely
spaced
$T_{c,n}$. These solutions all have s-wave symmetry and can  be thought of as oscillating excited states in an effective trapping potential which determines the hot region.
At $T_c = T_{c,0}$ only the non-oscillating solution $\Delta_0(\theta)$ develops, but other $\Delta_n (\theta)$ are induced by $\Delta_0 (\theta)$ and get resonantly enhanced below $T_{c,n}$.
Interference effects from these resonantly induced components modify the width of the hot region and make it temperature-dependent.

The existence of series of low energy states in the superconducting dome of quantum-critical systems raises several intriguing possibilities. These states may participate in subgap dynamics \cite{Bardasis1961,Chubukov1999} and enhance superconducting fluctuations.
Another possibility is the transition at some $T<T_c$ into a state with broken time-reversal symmetry~\cite{Sigrist1998},
when components $\Delta_n$ emerge with an additional phase $e^{i\phi}$ compared to $\Delta_0$, and  $\phi \neq 0,\pi$.
\begin{acknowledgments}
  We thank E. Berg, R. Fernandes, S. Kivelson, M. N. Gastiasoro, S. Lederer and Y. Schattner for stimulating discussions. This work was supported  by the NSF DMR-1523036.
  We thank the Minnesota Supercomputing Institute at the University of Minnesota for providing resources that assisted with this work.
\end{acknowledgments}

\bibliography{QCsuperconductivity}
\end{document}


\title{
  Supplemental material for "Multiple pairing states and temperature-dependent gap anisotropy for superconductivity near a nematic quantum-critical point'' by Avraham Klein, Yi-Ming Wu and Andrey Chubukov}
\maketitle
In these supplementary notes we describe our analytic calculations, both for the linearized and the nonlinear gap equation. In the final section we briefly describe our numerical procedure. In what follows we first consider the pairing at a QCP, and then away from it. For the latter case we introduce a finite  correlation length $\xi \gg \kf^{-1}$.

\section{The linearized gap equation at a QCP}
\label{sec:line-gap-equat}

 We consider a generic $d-$wave form-factor  $f(\theta)$ and focus on the region  $-\pi/2 <\theta < \pi/2$.

We use as an input  previous works \cite{Lederer2015,Klein2018c}, in which the  Eliashberg gap equation has been derived. It has the form (Eq. (4) of the main text)
\begin{align}
  \label{eq:gap-eliashberg}
  &Z(\theta, \w_n) \Delta(\theta,\w_n) = \nn\\
  & T \sum_{\w_m}\int \frac{d\phi}{\tp} \frac{\Delta(\theta+\phi,\w_m)f^2\left(\theta+\frac{\phi}{2}\right)}{|\w_m|}D(\theta,\phi,\w_m-\w_n),
\end{align}
where the bosonic propagator is
\begin{equation}
  \label{eq:D-def-1}
  D(\theta,\phi,\Omega) =
  \frac{\lambda}
  {\phi^2+ f^2\left(\theta+\frac{\phi}{2}\right)\frac{\lambda|\W|}{E_F\pi|\phi|}}.
\end{equation}
As we explained in the main text, the key effect of fermionic $Z (\theta\w_n)$  is to cancel the term with $\w_m=\w_n$ in the frequency sum in the r.h.s. of (\ref{eq:gap-eliashberg}).  We eliminate this term and thereafter set $Z (\theta, \w_n) =1$. This simplifies the analytical consideration. We keep the full $Z (\theta, \w_n)$ contribution in numerical calculations.

In order to make manifest the different roles played by the strong local fluctuations at the smallest angular transfers $\phi$  and the weak tail of the interaction at larger $\phi$,  it is convenient to split Eq. \eqref{eq:gap-eliashberg} into two parts. To do so, we subtract $\Delta (\theta, \w_m)$ from $\Delta(\theta + \phi,\w_m)$ in the numerator of the RHS of Eq. \eqref{eq:gap-eliashberg} and add
it 
 as a separate term.
  We then obtain
\begin{widetext}
\begin{equation}
  \label{eq:eff-eliash-1}
 \Delta(\theta,\w_n) = \hat\Lambda_0(\theta,\w_n)\Delta(\theta,\w_n) +
 T \sum_{\w_m \neq \w_n}\int \frac{d\phi}{2\pi} \frac{\left(\Delta(\theta+\phi,\w_m) -
     \Delta(\theta,\w_m)\right) f^2\left(\theta+\frac{\phi}{2}\right)}{|\w_m|}D(\theta,\phi,\w_m-\w_n).
\end{equation}
\end{widetext}
Here $\hat\Lambda_0$ takes care of 
strong local fluctuations,
\begin{eqnarray}
&& \hat\Lambda_0 \Delta(\theta,\w_n) =  T \sum \sum_{\w_m \neq \w_n} \frac{\Delta (\theta,\w_m)}{|\w_m|} \nonumber \\
&& \times \int \frac{d\phi}{2\pi} f^2\left(\theta + \frac{\phi}{2}\right) D(\theta,\phi,\w_m-\w_n).
\label{ex_11}
\end{eqnarray}
Because the integration over $\phi$ is confined to small $|\phi| \ll 1$, we can approximate $f(\theta + \phi/2)$ by $f(\theta)$,
and extend the limits of the $\phi$ integration to $\pm \infty$.  
 Integrating over $\phi$ we then obtain
\begin{align}
  \label{eq:Delta-strong}
  \hat\Lambda_0\Delta(\theta,\w_n) &\approx\frac{2}{3^{3/2}}T \sum_{m \neq n}\frac{\Delta(\theta,\w_m)}{|\w_m|}\frac{1}{\left(\frac{|\w_m-\w_n|}{\pi \lambda^2E_F f^4(\theta)}\right)^{1/3}}.
\end{align}
This is Eq. (5) of the main text.

The local gap equation
\begin{equation}
  \Delta(\theta,\w_n) = \hat\Lambda_0(\theta,\w_n)\Delta(\theta,\w_n)
  \label{ex_1}
\end{equation}
has a continuous set of solutions
\begin{equation}
  \label{eq:gap-loc-sol}
  \Delta (\theta,\w_n) = \Delta_\nu(\w_n) \delta(\theta - \nu)
\end{equation}
which progressively emerge at $T_c (\nu) = T_c (0) (1 - a_\nu \nu^2 + \ldots)$. The largest $T_c (0)$ is obtained by solving (\ref{ex_1})  with $f(\theta) = f(0)$. By order of magnitude, at a QCP, $T_c (\mu) \sim \lambda^2 E_F 
f^4 (\nu)$. 
To get the exact prefactor, we note the gap equation, Eq. (\ref{ex_1}),  with fermionic $Z$-factor re-introduced in the l.h.s.,  is equivalent to that for the quantum-critical $\gamma$ model with $\gamma =3$
(Ref. \cite{Wu2018a}). Using the results for the $\gamma$ model, we find 
 the
 exact expression
\begin{equation}
  T_c (0) = 0.022 \lambda^2 E_F
[f(0)]^{4}
\label{ex_2}
\end{equation}

The existence of a continuous set of solutions is a consequence of neglecting the second, non-local term in the r.h.s. of Eq. (\ref{eq:eff-eliash-1}). To obtain the actual gap function we need to account for this non-local term.
We proceed with an analytic treatment by making two approximations. First, we use the fact that the frequency sum converges, i.e., typical $\omega_m  = O(T)$. We then  neglect the  frequency dependence in the gap equation by taking $\w_n$ and $\w_m$ to be the smallest non-equal Matsubara frequencies, i.e., set $\omega_n= \pi T,\w_m = -\pi T$.  Second,  we expand $f (\theta)$ and $f (\theta + \phi/2)$  in small angles near $\theta=0, \phi=0$, i.e., around the center of the hot region. We assume and then verify that the expansion is justified everywhere in the hot region, since we will see that  the width of $\Delta (\theta, n)$, viewed as a function of $\theta$, is small, $\theta_h \sim \lambda^{1/3}$, as long as $\lambda \ll 1$.

Using these approximations,  we rewrite Eq. \eqref{eq:eff-eliash-1} as
\begin{widetext}
  \begin{equation}
    \label{eq:eff-gap-eq}
    \Delta (\theta) \left[1 - \left(1-a_\theta \theta^2\right)^{4/3}
      \left(\frac{T_c (0)}{T}\right)^{1/3}
    \right] = \frac{\lambda f^2 (0)}
    {\pi}
    \int \frac{d\phi}{2\pi} \frac{\Delta (\theta + \phi) - \Delta (\theta)}{\phi^2 + a_\lambda \frac{\lambda^3}{|\phi|}}
  \end{equation}
\end{widetext}
where we used $f(\theta) = f(0) (1 - a_\theta \theta^2)$, and $a_\theta$ and $a_\lambda$ are of order one.

A quick study of Eq. \eqref{eq:eff-gap-eq} shows that there are two relevant scales for $\theta$: a smaller (lower) scale $\theta_{l} = O(\lambda)$, below which
$\Delta (\theta) \approx \Delta (\theta =0)$ and a larger (higher) scale $\theta_h = O(\lambda^{1/3})$, which is set by balancing $\theta^2 \Delta (\theta)$ in the l.h.s. of
\eqref{eq:eff-gap-eq} and $ \lambda \int \frac{d\phi}{\pi} \frac{\Delta (\theta + \phi) - \Delta (\theta)}{\phi^2} \sim \lambda \Delta (\theta)/|\theta|$ in the r.h.s.  We will be interested in the gap function at $\theta \sim \theta_h >> \lambda$. In this region one can drop $\lambda^3/|\phi|$ term in the r.h.s. of  \eqref{eq:eff-gap-eq} and rewrite this equation as
\begin{widetext}
  \begin{equation}
    \label{eq:eff-gap-eq_1}
    \Delta (\theta) \left[1 - \left(\frac{T_c (0)}{T}\right)^{1/3}\right] =
     -\frac{4}{3} a_\theta \theta^2 \Delta (\theta) +
\frac{\lambda f^2 (0)}{\pi}
\int \frac{d\phi}{2\pi}\frac{\Delta (\theta + \phi) - \Delta (\theta)}{\phi^2}
  \end{equation}
\end{widetext}
Rescaling $\theta$ by $\theta_h$ and choosing the prefactor in $\theta_h \propto \lambda^{1/3}$ to eliminate the numerical factor between $\theta^2$ and integral terms in the r.h.s. of \eqref{eq:eff-gap-eq_1}, we re-write  \eqref{eq:eff-gap-eq_1} as
\newcommand{\btheta}{\bar\theta}
\newcommand{\bphi}{\bar\phi}
\begin{equation}
  \label{eq:gap-dimen-full}
  \eta  \Delta (\btheta) = \btheta^2 \Delta(\btheta) - \int_{-\infty}^{\infty}\frac {d\bphi}{\pi}\frac{\Delta(\btheta+\bphi)-\Delta(\btheta)}{\bphi^2}
\end{equation}
where  $\btheta,\bphi$ are rescaled angles
and we again extended the integration to $\pm \infty$.  The width of the ``hot" region in Eq. \eqref{eq:gap-dimen-full} (defined as  $|\theta| \leq \theta_h$)
 is of order one in the rescaled units. The parameter $\eta \sim ((T_c(0)/T)^{1/3}-1)/\lambda^{2/3}$. For $T \leq T_c (0)$, $\eta \sim (T_c (0) - T)/(T_{c}(0) \lambda^{2/3})$.

To solve Eq. \eqref{eq:gap-dimen-full} we treat separately the behavior in and out of the hot region, i.e., at $\btheta \ll 1$ and $\btheta \gg 1$. For $\btheta \gg 1$ we may neglect the $\Delta(\btheta)$ term in the integrand. The largest contribution to the integral comes from the hot region, i.e. from $\bphi \sim -\btheta$, with the width of order $\bphi = O(1)$, so to leading order we have
\begin{equation}
  \label{eq:Delta-out}
  \Delta (\btheta \gg 1) \approx  \frac{\bar \Delta}{\btheta^4},
\end{equation}
where
\begin{equation}
  \label{eq:Delta-h-deg}
  {\bar \Delta}  = \int \frac{dx}{\pi}  \Delta(x)
\end{equation}
In original variables, Eq. (\ref{eq:Delta-out})  shows that $\Delta (\theta)$ rapidly drops once $\theta$ exceeds $\theta_h$.
Inside the hot region, at $\btheta <1$, we can transform to Fourier components
$\Delta (\btheta) = (2\pi)^{-1} \int d \xi \exp{i \xi \btheta}  \Delta (\btheta)$ and reduce
Eq. \eqref{eq:gap-dimen-full}  to the Airy equation
\begin{equation}
  \label{eq:airy-supp}
  \eta \Delta (\xi) = -\pd_\xi^2\Delta(\xi)+|\xi|\Delta(\xi)
\end{equation}
This is Eq. (7) from the main text,
The boundary condition for the even gap function is $\Delta'(0)=0$. Another boundary condition is $\Delta (\xi \gg 1) \to 0$.
Using asymptotic expressions for the Airy functions we then obtain a discrete set of solutions, specified by integer numbers.
The solutions are
\begin{equation}
  \label{eq:Airy-sols}
\Delta_n(\xi)=\Ai(|\xi| - \eta_n),
\end{equation}
where $\eta_n$ is the "coordinate" of the $n^{\text{th}}$ zero
of the derivative of the  Airy function $\Ai'(|x|)$.
This implies that there exists a discrete set of solutions for the gap $\Delta_n (\xi)$ with $T_{c,n}$ set by  $A' (\eta_n) =0$.
The corresponding $\Delta_n (\btheta)$ changes sign $n$ times in the first octant ($8 n$ times over the whole Fermi surface).
For $n\gg 1$,
\begin{equation}
  \label{eq:eta-zeros}
  \frac{2}{3}|\eta_n|^{3/2} \approx \frac{\pi}{4}+n\pi,
\end{equation}
such that
\begin{equation}
  \label{eq:eta-n}
T_{c,n} = T_c (0) \left(1 - O(\lambda n)^{2/3}\right)
\end{equation}
The highest $T_c = T_{c,0}$ is for the solution with $n=0$. The corresponding $\Delta_0 (\btheta)$ does not change sign.

\section{Nonlinear gap equation}
\label{sec:nonl-gap-equat}

In this section we solve the nonlinear gap equation. We extend Eq. \eqref{eq:eff-eliash-1} by changing the frequency term in the numerator of the RHS to $|\omega_{n'}|\to \sqrt{\omega_{n'}^2+\Delta^2(\theta+\phi,\omega_{n'})}$
For $\Delta/T_c\ll 1$ it is enough expand to 3rd order in $\Delta$ and keep only the contribution to this order from the operator $\hat \Delta_0$ of Eq. \eqref{eq:eff-eliash-1}. We then repeat the steps to give us an effective equation for angles $\btheta,\bphi$ like Eq. \eqref{eq:gap-dimen-full}, but with additional terms that we will show below.

Equivalently, we can write a Ginzburg Landau Free energy and compute its saddle-point equations. We write the gap function as an expansion in the states $\Delta_n(\btheta)$ that are solutions of the linearized equation, $\Delta(\btheta) =\sum_n \bar\Delta_n\Delta_n(\btheta)$.
For convenience, we normalize them to be orthonormal and such that $\Delta_n(\btheta = 0)\propto (-1)^{n}$, i.e. nonoscillating $\Delta_0$ is positive for $\Delta(\btheta = 0)$, then $\Delta_1(\btheta=0)$ is negative, etc. The Free energy has the form,
\begin{equation}
  \label{eq:GL-q}
  F \propto \int -\frac{1}{2}\sum_n (t - t_n) \bar\Delta_n^2 + F_4 + \cdots
\end{equation}
where where $t_n = (T_c - T_{c,n})/T_c, t = (T_c - T)/T_c$,
\begin{equation}
  F_4 = A \sum_{ijkl} \bar\Delta_i\bar\Delta_j\bar\Delta_k\bar\Delta_l\int d\btheta \Delta_i(\btheta)\Delta_j(\btheta)\Delta_k(\btheta)\Delta_l(\btheta),
\end{equation}
and $A$ is a constant of order one. We simplify the analysis of  Eq. \eqref{eq:GL-q} by keeping only those cross-terms (terms with at least two different $\bar\Delta_n$) that have a power of $\bar\Delta_0$. The justification for this is that the numerical solution of the gap equation shows no oscillations, implying $\bar\Delta_0 \gg \bar\Delta_{n>0}$. Additionally, the numerical solution is real, indicating that the coefficients $\bar\Delta_n$ are real. The result is Eq. (9) of the main text, which we reproduce here,
\begin{align}
  \label{eq:F4-approx}
  F_4&\approx \sum_n\frac{a_{n}}{4}\bar{\Delta}_n^4 \nn\\
  &\qquad+ \sum_{n>0}\left[a_{n1} \bar\Delta_n^3\bar\Delta_0 + a_{n2}  \bar\Delta_0^2\bar\Delta_n^2 +a_{n3}\bar\Delta_0^3\bar\Delta_n
 \right].
\end{align}
Qualitatively, we expect that $|a_{3n}|,|a_{1n}| \ll |a_{2n}| \lesssim |a_n|$. This is because $a_{2n}$ is an integral over a positive-definite quantity whereas the integrals that determine $a_{3n},a_{1n}$ oscillate. In addition, we expect that $a_{3n} \propto (-1)^n$. This is because $\Delta_0^3(\btheta)$ is peaked in the region near $\btheta = 0$, and so the overall sign should go as the sign of $\Delta_n(\btheta = 0)$. Numerically, we find that the overlap integral in Eq. \eqref{eq:F4-approx} yields (up to the constant $A$),
\begin{align}
  \label{eq:a-coeffs}
  a_n &= 3.4, 2.6, 2.3, 2.15,\ldots & (n\geq 0) \nn\\
  a_{n1} &= -0.4, -0.1, -0.07 \ldots & (n > 0) \nn \\
  a_{n2} &= 2.0, 1.7, 1.5, \ldots & (n > 0) \nn\\
  a_{n3} &= -1.4, 0.2, -0.1, \ldots & (n > 0)
\end{align}
Since $a_{n1} \sim a_{n3}$ and $|\bar\Delta_n|\ll |\bar\Delta_0|$, we can neglect the $a_{1n}$ terms in $F_4$. We neglect the contribution of the induced states $\Delta_{n>0}$ on the nonoscillating $\Delta_{0}$, in which case the saddle point equation for $\bar\Delta_0$ immediately yields $\bar\Delta_0\approx\sqrt{t/a_0}$. The saddle-point solution of Eq. \eqref{eq:GL-q} for $n>0$ is
\begin{equation}
  \label{eq:bar-n-saddle}
  \bar\Delta_n \approx - \frac{a_{n3}\bar\Delta_0^3+a_n\Delta_n^3}{t_n + 2a_{n2}\bar\Delta_0^2-t}.
\end{equation}
The results quoted in the main text (after Eq. (9)) correspond to the limits $a_{n3}\bar\Delta_0^3 \gg  a_n \bar\Delta_n^3$ and $a_{n3}\bar\Delta_0^3 \ll  a_n \bar\Delta_n^3$. Note, that in the latter limit there are two possible solutions to the equation, with opposite signs. The correct sign is determined by the behavior for small $\bar\Delta_n$, i.e. by the sign of $-a_{n3}$.
\begin{figure*}[t!]
  \begin{minipage}{0.45\hsize}
  \centering
  \includegraphics[width=\hsize]{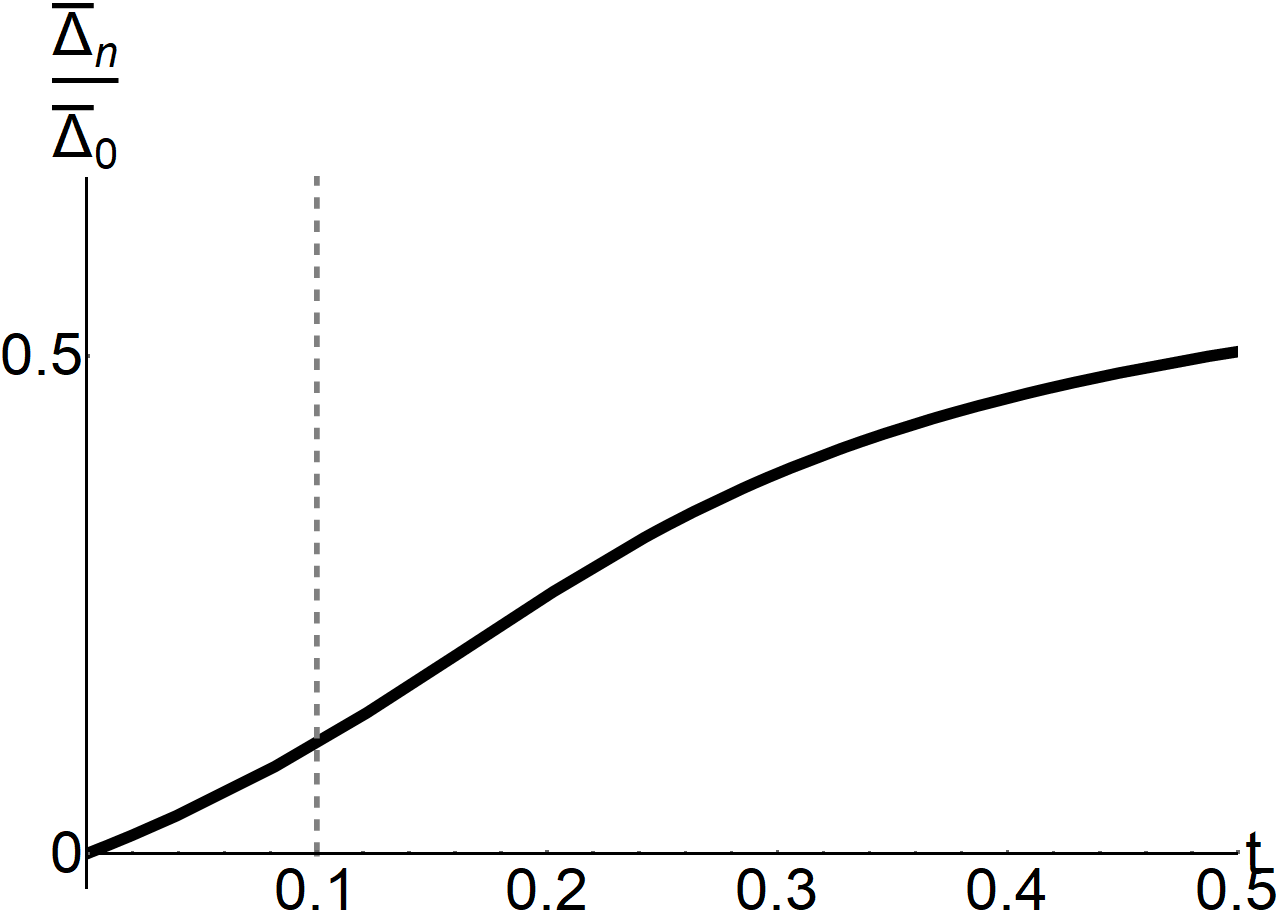}
  \caption{An illustration of the numerical solution of Eq. \eqref{eq:bar-n-saddle}. We chose the parameters $A = 1, t_n = 0.1, a_0 = 1, a_n = 0.7, a_{2n} = 0.4, a_{3n} = -0.03$.}
  \label{fig:bar-n-ill}
  \end{minipage}\hfill
  \begin{minipage}{0.45\hsize}
  \centering
  \includegraphics[width=\hsize]{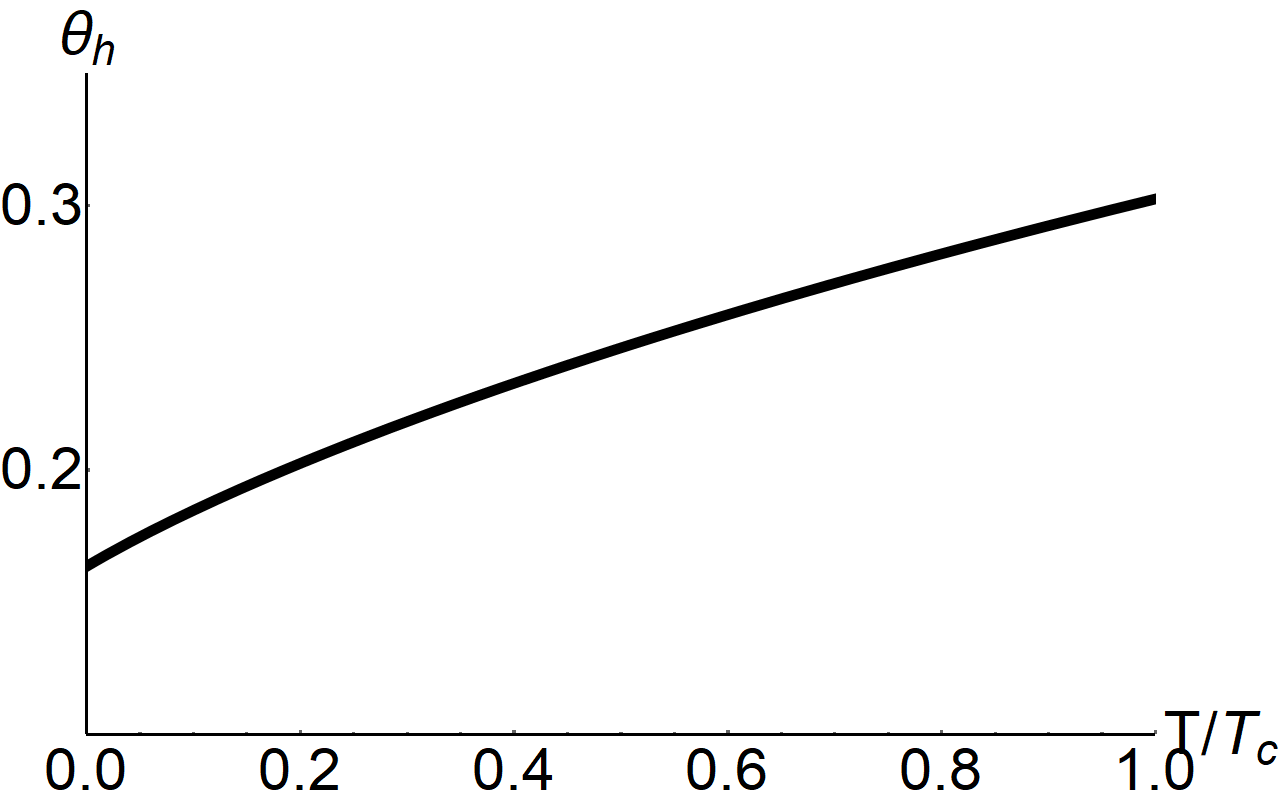}
  \caption{Illustration of the evolution of $\theta_h$ using Eq. \eqref{eq:theta-h}. For this figure we greatly simplified the expected dependence on the various $a_n,a_{nj}$ and used $A = 1, a_0 = a_n = 1, a_{n2} = 0.8, a_{n3} = 0.0125$. The numbers were chosen to avoid the artificial pole in Eq. \eqref{eq:theta-h}.}
  \label{fig:theta-h-ill}
  \end{minipage}
\end{figure*}
Fig. \ref{fig:bar-n-ill} depicts the numerical solution of Eq. \eqref{eq:bar-n-saddle}. Using the solution of Eq. \eqref{eq:bar-n-saddle} we can compute the $t$ dependence of the width of the hot region,
\begin{align}
  \label{eq:theta-h}
  \theta_h(t) &\propto \int d\bphi \frac{\Delta(\bphi)}{\Delta(\btheta=0)} \approx \frac{\sum\bar\Delta_n\int d\bphi\Delta_n(\bphi)}{\sum\bar\Delta_n\Delta_n(\btheta=0)} \nn\\
  &\approx \frac{\sum_n\bar\Delta_n\Delta_n(\xi=0)}{\sum_n\bar\Delta_n\Delta_n(\btheta=0)}
\end{align}
We can simplify Eq. \eqref{eq:theta-h} using the following analytic properties of Airy functions: $\Ai(-\xi_n) \propto (-1)^{n}/n^{1/4}, \int d\xi \Ai^2(|\xi|-\xi_n) \propto \sqrt{\xi_n} \sim n^{1/3}, \int d\xi \Ai(|\xi|-\xi_n) \approx \mbox{const}$. These in turn imply $\Delta(\btheta=0) \propto (-1)^n/n^{1/6}, \Delta(\xi=0) \propto 1/n^{1/6+1/4=5/12}$. Then we have
\begin{align}
  \label{eq:theta-h-2}
  \theta_h(t) &\propto \frac{1-a\sum_{n>0}\left|\frac{\bar\Delta_n}{\bar\Delta_0}\right|\frac{(-1)^{n+1}}{n^{5/12}}}
                {1-b\sum_{n>0}\left|\frac{\bar\Delta_n}{\bar\Delta_0}\right|\frac{1}{n^{1/6}}}\nn\\
  &\approx \frac{a'}{1-b'\sum_{n=1}^{N(t)}n^{-1/6}-b'' t\sum_{n>N(t)}t_n^{-1}n^{-1/6}}
\end{align}
where $N(t)\sim t^{3/2}$ is the largest $n$ such that $t > t_n$, and $a,b,a',b',b''$ are constants. Eq. \eqref{eq:theta-h-2} has an artificial pole at larger $t$, that arises from the many neglected terms in $F_4$, Eq. \eqref{eq:F4-approx}. For $t \ll 1$ we can expand to obtain
\begin{equation}
  \label{eq:theta-h-3}
  \theta_h(t) \propto 1 + c t + c' t^{5/4},
\end{equation}
in agreement with Eq. (10) of the main text.

\section{Away from a critical point}

The analysis of the discrete set of solutions of the gap equation can be readily extended to a finite nematic correlation length $\xi$. The bosonic propagator at a finite $\xi$ is
\begin{equation}
  \label{eq:D-def-1_1}
  D(\theta,\phi,\Omega) =
  \frac{\lambda}
  {\phi^2+ (k_F \xi)^{-2} + f^2\left(\theta+\frac{\phi}{2}\right)\frac{\lambda|\W|}{E_F\pi|\phi|}}.
\end{equation}
The computational steps are the same as before. The local gap equation is  given by (\ref{ex_1}) and still have an infinite number of solutions.
To study the effect of 
  a 
  finite correlation length, it is useful to consider first the perturbative regime $(\kf\xi)^{-1} \ll \lambda$ and then go to the opposite regime $(\kf\xi)^{-1} \gg \lambda$.
The operator $\hat\Lambda_0(\theta, \w_n)$ is given by
(\ref{ex_11}).  For $(k_F \xi)^{-1} \ll \lambda$
 we have 
\begin{flalign}
  \label{eq:delta-strong-finite}
  \hat\Lambda_0(\theta) \Delta(\theta,n) \approx \nn\hspace{150pt}\\ \left(\frac{T_c(0)}{T}\right)^{1/3}
  \left(1 - a_\xi \frac{(\kf\xi)^{-2}}{\lambda^2}\right)
  \Delta(\theta,n) \nn\\
\qquad\qquad  -a_\theta\theta^2\Delta(\theta,n)
\end{flalign}
where $a_\xi = O(1)$.
 In the nonlocal part of the equation, the finite $\xi$ plays no role since the behavior at small $\phi$ is smooth. Consequently, the gap equation with the nonlocal term has the same form as Eq. \eqref{eq:gap-dimen-full}, but with a new scaling for $\eta$,
\begin{equation}
  \label{eq:eta-scale-pert}
  \eta \lambda^{2/3} \propto \left(\frac{T_{c} (0)}{T}\right)^{1/3}  
  \left(1 - a_\xi \frac{(\kf\xi)^{-2}}{\lambda^2}\right)
   -1.
\end{equation}
The new onset temperatures are,
\begin{align}
  \label{eq:Tc-n-pert}
  T_{c,n} &\approx T_{c,0}\left[
\frac{1 - a_\xi \frac{(\kf\xi)^{-2}}{\lambda^2}}  
  {1+b(\lambda n)^{2/3}}\right]^3 \nn\\
          & \approx T_c(0) \left(1 - \left(\frac{\bar{k}_F\xi}{\lambda}\right)^2\right)(1-(\bar{\lambda}n)^{2/3}),
\end{align}
where $\bar{k}_F\sim \kf$ and $\bar{\lambda} \sim \lambda$. We see that,
 at $(k_F \xi)^{-1} \ll \lambda$,  the effect of the finite correlation length is a uniform reduction of all onset temperatures,
   but multiple solutions survive and the width of the hot region $\theta_h$ does not depend on $\xi$. 

In the opposite limit $(k_F \xi)^{-1} \gg \lambda$, $\Lambda_0$ has a BCS form,
\begin{align}
  &\hat\Lambda_0(\theta,\w_n) \Delta (\theta, \w_n) \approx  \frac{\lambda f^2 (\theta)}{2 (k_F \xi)^{-1}} T\sum_m^{\left\lfloor\frac{E_F(\kf\xi)^{-3}}{\lambda T}\right\rfloor} \frac{\Delta (\theta, \w_m)}{|\w_m|} \nn\\
  &~~\approx \frac{\lambda}{2(\kf\xi)^{-1}}\log\left(\frac{E_F(\kf\xi)^{-3}}{\lambda T}\right)\Delta(\theta,\w_n) - a_\theta\theta^2\Delta(\theta,\w_n)
\label{ex_22}
\end{align}
Eq. \eqref{ex_22} shows that the frequency sum no is no longer limited to a few first Matsubara frequencies, and instead gives a logarithm and determines an onset temperature $\bar{T}_c(0)\sim (E_F(\kf\xi)^{-3}/\lambda)\exp\left(-(\kf\xi)^{-1}/\lambda\right)$.
This logarithm then appears in the nonlocal part of the gap equation as well, and affects the width of the hotspot $\theta_h$.
The gap equation with the non-local term 
becomes
  \begin{align}
    \label{eq:ex_3}
    &\Delta(\theta)\frac{\bar T_c(0)-T}{\bar T_c(0)}\frac{\lambda}{2(\kf\xi)^{-1}} \propto\nn\\
    &\qquad\quad-a_\theta\theta^2 + (\kf\xi)^{-1}\int d\phi \frac{\Delta(\theta+\phi)-\Delta(\theta)}{\phi^2},
  \end{align}
Eq. \eqref{eq:ex_3} can be recast into the same form as the dimensionless Eq. \eqref{eq:gap-dimen-full}, but now
\begin{align}
  \label{eq:scaling-finite}
  \theta_h &\propto (\kf\xi)^{-1/3}
\end{align}
and
\begin{align}
  \label{eq:eta-scaling-finite}
  \eta \propto \frac{\bar T_c-T}{\bar T_c}\frac{\lambda}{(\kf\xi)^{-5/3}}
\end{align}
Eq. \eqref{eq:eta-scaling-finite} implies that in order for $T_{c,n}$ to be close in temperature the system must obey
\begin{equation}
  \label{eq:QC-condition}
  \frac{(\kf\xi)^{-5/3}}{\lambda}\ll 1.
\end{equation}
Eq. \eqref{eq:QC-condition} demonstrates that the strong evolution of $\theta_h$ with temperature is a signature of quantum-critical pairing. When $(\kf\xi)^{-1} \sim \lambda^{3/5}$, the solutions with $n>0$ do not develop at finite $T/T_c$, and the superconductivity is of a conventional nature, although it remains strongly anisotropic till $\kf\xi \sim 1$.

\section{Details of the numerical solutions}
\label{sec:deta-numer-solut}

We solved both the linearized and 
 the 
 nonlinear gap equation numerically. The details of the numerical solution of the linearized equation have appeared previously in Ref. \onlinecite{Klein2018c}. All the figures and numerical values that appear in the main text and this supplementary material were obtained for a value of $\lambda = 0.025$, with a numerical mesh of $512$ points in the range $-\pi/2 < \theta <\pi/2$, and 48 Matsubara frequencies (half negative and half positive).
Regarding the nonlinear gap equation, all the figures and numerical values that appear in the main text and this supplementary material were for a value of $\lambda = 0.03$. They were obtained with a mesh of 1000 points in the range $-\pi < \theta < \pi$ and 101 Matsubara frequencies, namely we take $\omega_n=(2n+1)\pi T$ with $n$ ranging from -50 to 50. The numerical solution was obtained by iteration. Both linear and nonlinear equations were solved in MATLAB (various versions).

To produce the figures in this supplementary material, as well as Fig. 4 in the main text, we employed Mathematica 11 and its implementation of the Airy function, with $\lambda = 0.03$.
\bibliography{QCsuperconductivity}

%% file: kleinPreamble.tex
\usepackage{amsmath}
\usepackage{epsf}
\usepackage{graphicx}

\usepackage{dcolumn}
\usepackage{bm}

\usepackage{amssymb}
\usepackage{array}
\usepackage{varioref}
\usepackage{wrapfig}
\usepackage{etoolbox}
\usepackage{color}

\providecommand{\nn}{\nonumber}

\providecommand{\pd}{\partial}

\providecommand{\bv}[1]{\bm{\mathrm{#1}}}

\providecommand{\w}{\omega}
\providecommand{\W}{\Omega}
\providecommand{\q}{\bv{q}}

\renewcommand{\k}{\bv{k}}

\providecommand{\ef}{\varepsilon_F}

\providecommand{\kf}{k_F}

\providecommand{\kf}{k_F}

\renewcommand{\q}{\bv{q}}

\providecommand{\tp}{2\pi}
\providecommand{\tpp}{\left(2\pi\right)}

